\documentclass{iaus} 
\usepackage{graphicx} 
\title[Supermassive Black Holes in BCGs] 
{Supermassive Black Holes in BCGs}
\author[Dalla Bont\`a et al.]   
{E. Dalla Bont\`a$^{1,2}$, L. Ferrarese$^2$, J. Miralda-Escud\'e$^3$,
 L. Coccato$^4$,\\ E. M. Corsini$^1$, A. Pizzella$^1$}
\affiliation{$^1$Dipartimento di Astronomia, Universit\`a di Padova, 
Padova, Italy\\[\affilskip] 
$^2$Herzberg Institute of Astrophysics, Victoria,
Canada\\[\affilskip]
$^3$Institut de Ciencies de l'Espai (CSIC-IEEC)/ICREA, Barcelona, Spain\\[\affilskip] 
$^4$Kapteyn Astronomical Institute, University of Groningen, Groningen, 
  The Netherlands\\[\affilskip]}
\pubyear{2006} 
\volume{238}  
\pagerange{001--999} 
\date{??? and in revised form ???} 
\jname{Black Holes: from Stars to Galaxies -- across the Range of Masses} 
\editors{V. Karas \& G. Matt, eds.} 
\begin{document} 
\maketitle 
\begin{abstract} 
We observed a sample of three Brightest Cluster Galaxies (BCGs), 
Abell~1836-BCG, Abell~2052-BCG, and Abell~3565-BCG, with the Advanced Camera for
Surveys (ACS) and the Imaging Spectrograph (STIS) on board the Space
Telescope. For each target galaxy we obtained high-resolution
spectroscopy of the H$\alpha$ and [N~{\small II}]$ \,\lambda6583$
emission lines at three slit positions, to measure the central
ionized-gas kinematics.
ACS images in three different filters (F435W, F625W, and
FR656N) have been used to determine the optical depth of the dust, 
stellar mass distribution near the nucleus, and intensity map. We
present supermassive black hole (SBH) mass estimates for two galaxies
which show regular rotation curves and strong central velocity
gradients, and an upper limit on the SBH mass of the third one.  For the
SBHs of Abell~1836-BCG and Abell~3565-BCG, we derived 
$M_\bullet=4.8^{+0.8}_{-0.7}\times 10^9 $ M$_\odot$ and 
$M_\bullet=1.3^{+0.3}_{-0.4}\times 10^9 $ M$_\odot$ at
$1\sigma$ confidence level, respectively. For the SBH of Abell~2052-BCG, we found
$M_\bullet \leq 7.3 \times10^9 $ M$_\odot$.

\end{abstract} 
 
\firstsection 
\section{Introduction} 
The link between the evolution of SBHs
and the hierarchical build-up of galaxies is imprinted in a number of
scaling relations connecting SBH masses to the global properties of
the host galaxies. The high SBH mass end of these relations has yet
to be fully explored: the massive galaxies expected to host the most 
massive SBHs are generally at large distances, making a dynamical
detection of SBHs observationally challenging.
This is unfortunate, since black holes with mass in excess of $10^9$ M$_\odot$ occupy an
integral part in our understanding of the co-evolution of SBHs and
galaxies: these are the systems that have undergone the most extensive and protracted history
of merging; moreover, they represent
the local relicts of the high redshift quasars detected in optical
surveys. To investigate the high-mass end of the SBH mass function, we 
selected three BCGs from the sample of Laine et
al. (2003). Their large masses, luminosities and stellar velocity
dispersions, as well as their having a merging history which is
unmatched by galaxies in less crowded environments, make these
galaxies the most promising hosts of the most massive SBHs in the
local Universe.

\section{Dynamical Modeling and Results}
Following Coccato et al. (2006), a model of the gas velocity field for
each galaxy
is generated by assuming that the ionized-gas component is moving onto
circular orbits in an infinitesimally thin disc centered at the galactic
nucleus. The model is projected onto the plane of the sky
for a grid of assumed inclination angles of the gaseous disk.  Finally, the
model is brought to the observational plane by accounting for the
width and location (namely position angle and offset with respect
to the disk center) of each slit, for the point spread function of the STIS
instrument, and for the effects of charge bleeding between
adjacent CCD pixels.  The mass of the SBH is determined by finding the
model parameters (SBH mass, inclination of the gas disk, and
mass-to-light ratio of the stellar component) that produce the best match to the observed velocity
curves. 

For Abell~1836-BCG, we derive $M_\bullet=4.8^{+0.8}_{-0.7}\times 10^9$ M$_\odot$, the largest SBH mass to have been
dynamically measured to-date. The best fit inclination angle is
$i=76^\circ\pm 1^\circ$,  while only an upper limit to the stellar
mass-to-light ratio is found ($M/L_I \leq 4.0$ M/L$_I$$_\odot$ at
$1\sigma$ confidence level). For Abell~3565-BCG we determine
$M_\bullet=1.3^{+0.3}_{-0.4}\times 10^9$ M$_\odot$, with
$i=50^\circ\pm 1^\circ$ and $M/L_I =9.0\pm0.8 $ M/L$_I$$_\odot$ 
($1\sigma$ confidence level).  In the case of Abell~2052-BCG we find
$M_\bullet\leq 7.3 \times 10^9 $ M$_\odot$, following the method by
Sarzi et al. (2002).

In Fig. 1 we show the location of our SBH masses determinations 
in the near-infrared $M_\bullet-L_{bulge}$ relation of
Marconi \& Hunt (2003) and the $M_\bullet-\sigma_c$ relation,
as given in Ferrarese \& Ford (2005). Implications of these
observations will be discussed in a forthcoming paper (Dalla Bont\`a
et al., in preparation).
\begin{figure}
\begin{center}
 \includegraphics[height=7cm]{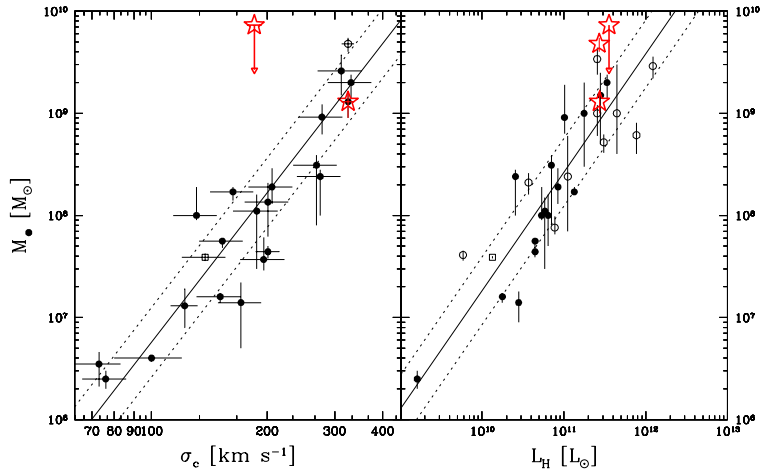}
  \caption{Location of the SBHs masses of our BCG sample galaxies
    (star symbols) with respect to the $M_\bullet-\sigma_c$ relation
    of Ferrarese \& Ford (2005, left panel, $\sigma_c$ for the BCGs
    are from Smith et
    al. 2000 ) and near-infrared $M_\bullet-L_{bulge}$ relation of
    Marconi \& Hunt (2003, right panel). In the left panel, following
    Ferrarese \& Ford (2005) we plot the SBH masses based on resolved
    dynamical studies of ionized gas (open circles), water masers
    (open squares), and stars (filled circles). In the right panel we
    plot the SBH masses for which $H-$band luminosity of the host
    spheroid is available from Marconi \& Hunt (2003).  In both panels,
    dotted lines represent the $1\sigma$ scatter in $M_\bullet$. No
    $\sigma_c$ is yet available for Abell~1836-BCG.}
\end{center}
\end{figure}
 

\begin{thebibliography}{} 
\bibitem[Coccato et al.(2006)]{2006MNRAS.366.1050C} 
  {Coccato, L., et al.} 2006,
  \textit{MNRAS}, 366, 1050
\bibitem[Ferrarese \& Ford(2005)]{2005SSRv..116..523F} 
  {Ferrarese, L., \& Ford, H.} 2005, \textit{Sp. Sci. Rev.}, 
  116, 523 
\bibitem[Laine et al.(2003)]{laine03} {Laine, S., et al.} 2003, 
\textit{ApJ}, 125, 478
\bibitem[Marconi \& Hunt(2003)]{marhunt03} Marconi, A., \& Hunt, 
L.~K.\ 2003, \textit{ApJ}, 589, L21 
\bibitem[Sarzi et al.(2002)]{2002ApJ...567..237S} 
  {Sarzi, M., et al.} 2002,
  \textit{ApJ}, 567, 237
\bibitem[Smith et al.(2000)]{smith00} {Smith, R.~J., et al.} 2000,
\textit{MNRAS}, 313, 469
\end{thebibliography}
\end{document}